\documentclass[pra,aps,showpacs,floatfix,twocolumn]{revtex4-2}

\usepackage{amsmath}
\usepackage{amsfonts}
\usepackage{amssymb}
\usepackage{revsymb}
\usepackage{graphicx}
\usepackage{physics}
\usepackage{xcolor}
\usepackage{booktabs}
\usepackage{tabularx}
\usepackage{multirow}
\usepackage{url}
\graphicspath{{figs/}}

\usepackage{dcolumn}
\usepackage{bm}
\usepackage{braket}
\usepackage{float}

\newcommand{\Lu}{$^{176}\mathrm{Lu}^+$}

\begin{document}

\preprint{APS/123-QED}

\title{Zeeman Degenerate Sideband Cooling in $\mathbf{^{176}}$Lu$\mathbf{^+}$}
\author{Qin Qichen}
\affiliation{Centre for Quantum Technologies, National University of Singapore, 3 Science Drive 2, 117543 Singapore}
\author{Qi Zhao}
\affiliation{Centre for Quantum Technologies, National University of Singapore, 3 Science Drive 2, 117543 Singapore}
\author{M. D. K. Lee}
\affiliation{Centre for Quantum Technologies, National University of Singapore, 3 Science Drive 2, 117543 Singapore}
\author{Zhao Zhang}
\affiliation{Centre for Quantum Technologies, National University of Singapore, 3 Science Drive 2, 117543 Singapore}
\author{N. Jayjong}
\affiliation{Centre for Quantum Technologies, National University of Singapore, 3 Science Drive 2, 117543 Singapore}
\author{K. J. Arnold}
\affiliation{Centre for Quantum Technologies, National University of Singapore, 3 Science Drive 2, 117543 Singapore}
\affiliation{Temasek Laboratories, National University of Singapore, 5A Engineering Drive 1, Singapore 117411, Singapore}
\author{M. D. Barrett}
\affiliation{Centre for Quantum Technologies, National University of Singapore, 3 Science Drive 2, 117543 Singapore}
\affiliation{Department of Physics, National University of Singapore, 2 Science Drive 3, 117551 Singapore}
\email{phybmd@nus.edu.sg}


\begin{abstract}
We explore degenerate Raman sideband cooling in which neighboring Zeeman states of a fixed hyperfine level are coupled via a two-photon Raman transition.  The degenerate coupling between $\ket{F,m_F}\rightarrow \ket{F,m_F-1}$ facilitates the removal of multiple motional quanta in a single cycle. This method greatly reduces the number of cooling cycles required to reach the ground state compared to traditional sideband cooling. We show that near ground state cooling can be achieved with a pulse number as low as $\bar{n}$ where $\bar{n}$ is the average phonon number in the initial thermal state.  We demonstrate proof-of-concept in $^{176}\mathrm{Lu}^+$ by coupling neighboring Zeeman levels on the motional sideband for the $F=7$ hyperfine level in $^3D_1$. Starting from a thermal distribution with an average phonon number of 6, we demonstrate near ground-state cooling with $\sim10$ pulses.  A theoretical description is given that applies to any $F$ level and demonstrates how effective this approach can be.
\end{abstract}

\maketitle

\section{\label{sec:level1}Introduction}

Trapped ions represent one of the most precisely controllable quantum systems, offering exceptional isolation from the environment and long coherence times. These characteristics have enabled their use in a wide range of forefront applications, including precision spectroscopy, quantum logic clocks, quantum simulation, and quantum information processing~\cite{ludlow2015optical,zhiqiang2023176lu,huntemann2016single,herschbach2012linear,blatt2012quantum,blatt2008entangled,haffner2008quantum}. In such systems, the ion's motion in the trapping potential is well described as a quantized harmonic oscillator, with discrete motional states $\ket{n}$. For many applications, particularly those involving entangling gates or quantum-logic spectroscopy, it is essential to initialize the ion in its motional ground state ($n = 0$) to ensure high-fidelity control and minimal decoherence~\cite{schmidt2005spectroscopy,wineland2013nobel}.

Doppler cooling serves as the initial stage in ion preparation, reducing the ion’s thermal energy to the millikelvin regime. To further cool to the motional ground state, resolved sideband cooling (SC) ~\cite{itano1995cooling} is often employed \cite{monroe1995resolved,hemmerling2011single,burd2016vecsel,deslauriers2004zero,goodwin2016sideband}.  In this technique a so-called red-sideband transition coherently transfers population from $\ket{\downarrow,n}$ to $\ket{\uparrow,n-1}$, where $\uparrow,\downarrow$ denote two internal states of the ion and $n$ the vibrational state.  The internal state is then reset via optical pumping to $\ket{\downarrow,n-1}$ provided the recoil energy from the scattering process is less than the vibrational energy spacing.  Repeated cycles drive the ion to the ground state.  

While traditional SC is highly effective for systems with a simple internal structure, it becomes less so for atoms with large nuclear spin. With a large number of available ground states, the optical repumping step requires a large number of spontaneous scattering events to reset the internal state, which increases the required optical pumping time, and the associated recoil heating diminishes cooling efficiency.  To mitigate these limitations, we employ degenerate Raman sideband cooling (DRSC) in which neighboring Zeeman states of a fixed hyperfine level are coupled via a two-photon Raman transition. The degenerate coupling of $\ket{F,m_F}\rightarrow \ket{F,m_F-1}$ transfers population across the entire hyperfine manifold.  When driven on the red sideband, multiple vibrational quanta are removed in a single pulse.

In this work, we demonstrate DRSC in $^{176}$Lu$^+$, which has a nuclear spin $I = 7$ giving a total of 45 Zeeman sublevels within the $^3D_1$ manifold. Cooling is implemented by coherently coupling the state $\ket{^3D_1, F = 7, m_F = 0}$ to $\ket{^3D_1, F = 7, m_F = -7}$ via a sequence of Zeeman-degenerate red sideband transitions.  Starting from a Doppler-cooled thermal state with an average motional occupation of $\bar{n}=6.9(1.1)$ after Doppler cooling, $\bar{n}$ is reduced to $0.118(14)$ using just 10 DRSC pulses.  For comparison, standard first-order SC protocols under similar conditions would typically require approximately 50 cycles, assuming idealized pulse dynamics~\cite{optimize}.  We also demonstrate a conditional state preparation technique, which we call Raman dark preparation (RDP), that further reduces the occupation to $\bar{n} = 0.0129(67)$.  This conditional step is qualitatively similar to ``erasure correction cooling" reported in \cite{shaw2025erasure}.  Our results establish DRSC as a powerful and efficient technique for preparing ions in the motional ground state.

Section~II presents simulations and a theoretical description that illustrates the advantages of the method.  Our demonstration on the $F=7$ manifold is motivated by our use of the $m=0$ states for clock applications.  However, we also give a theoretical analysis for cooling on the $F=8$ as this provides more efficient cooling and may be relevant to other systems.  Section~III describes the experimental setup and key calibration procedures for the implementation. The main experimental results are presented in Section~IV, followed by a discussion of limitations and outlook in Section~V.

\section{Degenerate Raman Sideband Cooling}

After Doppler cooling, the ion’s motion along each trap axis is well approximated by a thermal distribution given by
\begin{equation}
p_{\bar{n}}(n) = \frac{\bar{n}^n}{(\bar{n} + 1)^{n+1}}.
\end{equation}
where $\bar{n}$ is the mean occupation.  The standard Doppler cooling limit on a transition with linewidth $\Gamma$ is $\bar{n} \sim \Gamma/(2\omega)$,
where $\omega$ is the trap oscillation frequency~\cite{stenholm1986semiclassical,eschner2003laser}. 

When coupling to the motion, the Rabi frequency for driving transitions between Fock states $\ket{n}$ and $\ket{n'}$ is given by~\cite{wineland1979laser,wineland1998experimental}:
\begin{equation}
\label{DebyeWaller}
\Omega_{n,n'} = \Omega e^{-\eta^2/2} \sqrt{\frac{n_<!}{n_>!}} \eta^{|n - n'|} \mathcal{L}_{n<}^{|n - n'|}(\eta^2),
\end{equation}
where $\eta$ is the Lamb–Dicke parameter, $\mathcal{L}_n^{(\alpha)}$ is the generalized Laguerre polynomial, and $\Omega$ is an overall scale factor dependent on the laser intensity. Here, $n_>$ ($n_<$) denote the larger (smaller) of $n$ and $n'$. For cooling, it is typically sufficient to use the first-order sideband transitions ($|n - n'| = 1$), although it is sometimes necessary to use the second sideband when operating outside the Lamb-Dicke regime due to a vanishing of the first sideband coupling near $\eta\sqrt{n}\approx 1.9$ \cite{wan2015efficient,che2017efficient}.   In the Lamb-Dicke regime, $\eta\sqrt{\bar{n}}\ll 1$ and $\Omega_{n,n-1} \simeq \Omega \eta \sqrt{n}$.  A useful approximation for Eq.~\ref{DebyeWaller}, at least for the first sideband and the carrier, is given by
\begin{equation}
\label{DWApprox}
\Omega_{n,n+\Delta n}\approx \Omega J_{\Delta n}\left(2\eta\sqrt{n+\tfrac{1}{2}(\Delta n+1)}\right)
\end{equation}
where $J_k(x)$ is the Bessel function of the first kind.

\begin{figure}[htbp]
\begin{center}
\includegraphics[width=0.9\linewidth]{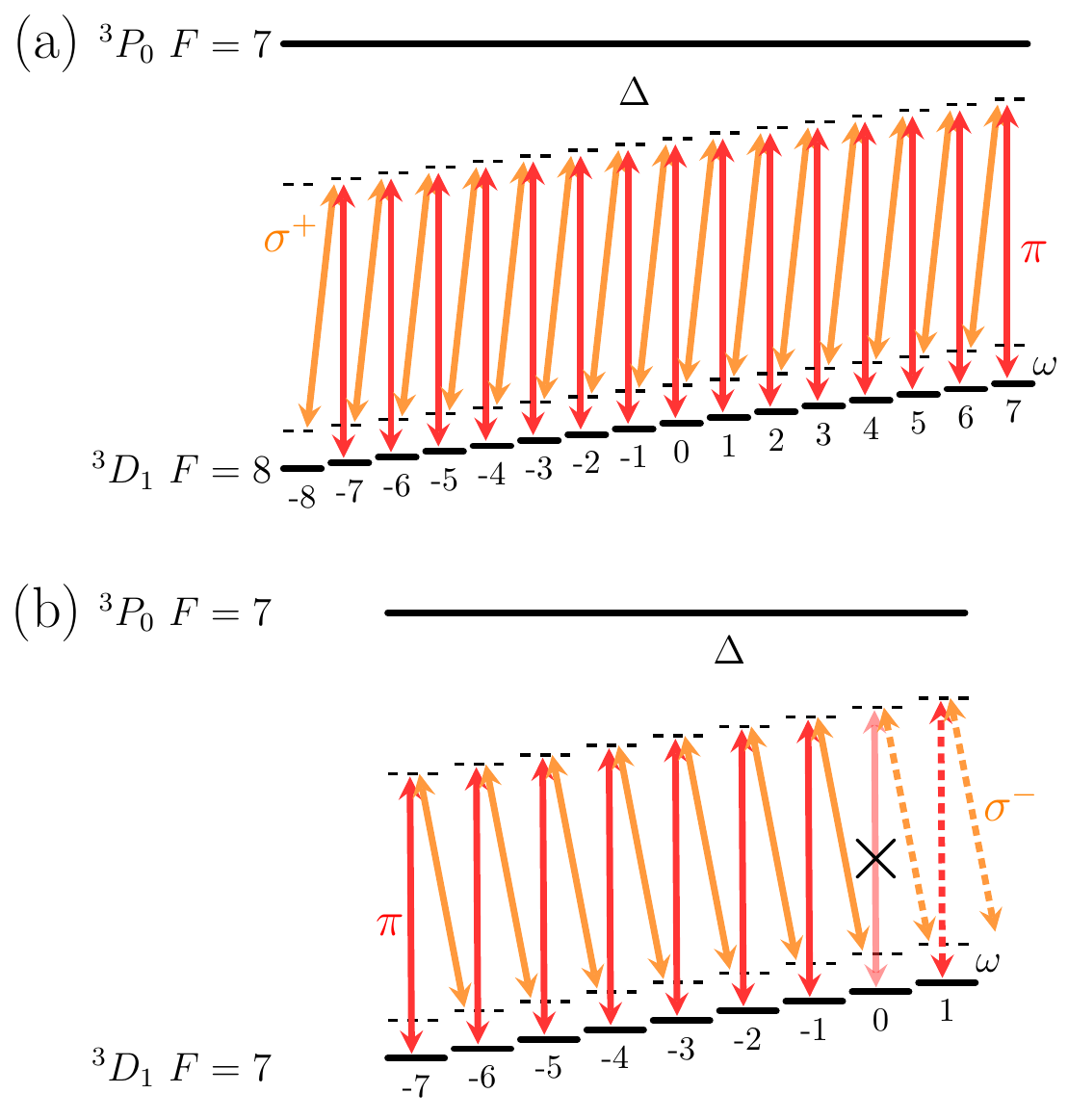}
\caption{\label{fig:Figure1} Schematic of Raman transitions for $F=8$ (a) and $F=7$ (b).  $\pi$ (red) and either $\sigma^+$ or $\sigma^-$ (orange) polarized fields couple adjacent $m_F$ states while lowering $n$.
}
\end{center}
\end{figure}

In principle, DRSC can be implemented using the Zeeman sublevels of any of the $F=6,7,8$ hyperfine manifolds in the $^3D_1$ state, and we illustrate two such possibilities in Fig.~\ref{fig:Figure1}. The Raman beams couple $\ket{F,m_F}$ to $\ket{F,m_F-1}$ using $\pi$ and $\sigma^+$ polarizations for $F=8$, and $\pi$ and $\sigma^-$ polarizations for $F=7$. For $F=8$, this configuration couples states across the entire manifold, but for the $F=7$ case, we are restricted to $m_F\leq0$ due to selection rules that prevent coupling to $m_F=1$.  In our system $\ket{8,\pm8}$ states are less suitable for clock interrogation, thermometry, or state-selective detection, and so the experimental demonstration utilizes the $F=7$ configuration.  However, we include numerical results for the $F=8$ case as it presents advantages for DRSC which may be applicable in other systems.

Starting from $\ket{F=7,m_F=0,n}$, Raman coupling between $\ket{F=7,m_F,n}$ and $\ket{7,m_F-1,n-1}$, drives population towards $m_F=-7$ removing one vibrational quanta for each decrease in $m_F$ allowing a maximum of 7 quanta to be removed in a single pulse.   In practice, ac Stark shifts from the Raman beams and quadratic Zeeman shifts from the applied static magnetic field will move the system away from degeneracy.  However, setting the intensities such that $I_{\sigma^-}/I_\pi = 2$ cancels the contribution from the tensor polarizability leaving only the vector contribution from the $\sigma^-$ beam, which is equivalent to an effective magnetic field along the same direction as the applied field~\cite{le2013dynamical}.  Moreover, the tensor polarizability has the same form as the quadratic Zeeman shift, so this can also be tuned to zero by a small mismatch in the  $I_{\sigma^-}/I_\pi = 2$ condition.  For lutetium this is unnecessary owing to the large hyperfine splittings and hence negligible quadratic Zeeman shift, but degeneracy can be assured in general.

In choosing the polarization care should be taken when using a level structure that supports a tensor polarizability with a small Zeeman splitting as is the case for \Lu.  When the level supports a tensor polarizability, components of the polarization from a single beam can induce off-resonant Raman couplings between Zeeman states, which vanish in the absence of a tensor polarizability.  When the Zeeman splitting is small relative to the Raman coupling strengths, it can result in an unwanted precession between Zeeman states.  For this reason, we have chosen a configuration in which each beam polarization has only a single component in the spherical basis, which avoids any and all unwanted Raman couplings.

\begin{figure}[h]
    \centering
   \includegraphics[width=1.0\linewidth]{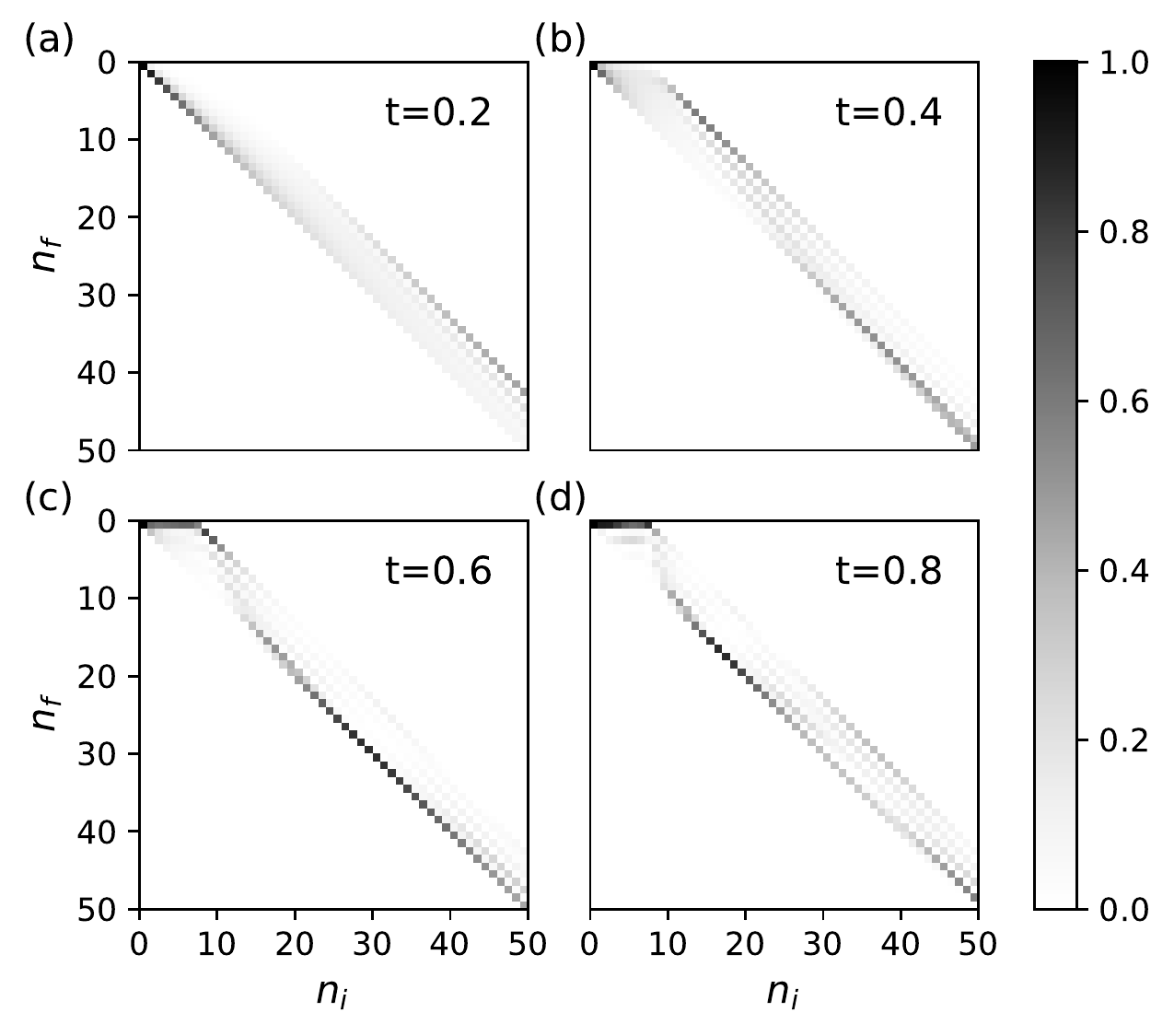}
    \caption{
        \label{fig:Wmatrix}Graphical representation of the first $50\times50$ elements of the population transfer matrix. (a) through (d) depict the population transfer matrices for pulse durations $t = [0.2,\, 0.4, \,0.6 ,\,0.8] \times T_f$.
    }
\end{figure}

As the Raman transfer is followed by incoherent repumping to $\ket{F=7,m_F=0,n}$, any coherence built up during the Raman pulse is irrelevant and we consider only the vibrational population dynamics which can be represented as a matrix transformation~\cite{optimize}. The initial thermal distribution after Doppler cooling is encoded in the vector $\vec{p}_{\bar{n}_i} = \{p_{\bar{n}_i}(0), \dots, p_{\bar{n}_i}(n_{\max})\}$ with $n_\mathrm{max}$ chosen to include the desired fraction of the population.  The effect of a single Raman pulse is then described by a population transition matrix $W(t)$, where each element $a_{ij}(t)$ represents the probability of population transfer from state $n = i$ to $n = j$ for a pulse duration $t$.  For the $F=7$ case under consideration, population may transfer seven steps at most, resulting in $W(t)$ being an upper triangular banded matrix with a bandwidth of eight.  Explicitly, it has the form
\begin{equation*}
    W(t) =
\begin{pmatrix}
1 & a_{10}(t) & a_{20}(t) & \cdots & a_{70}(t) & 0 & 0 & \cdots\\
0 & a_{11}(t) & a_{21}(t) & \cdots & a_{71}(t) & a_{81}(t) & 0 & \cdots\\
0 & 0 & a_{22}(t) & \cdots & a_{72}(t) & a_{82}(t) & a_{92}(t) & \cdots\\
\vdots & \vdots & \vdots & \vdots & \vdots & \vdots & \vdots & \ddots\\
\end{pmatrix}.
\end{equation*}
After a pulse sequence $\{t_0, t_1, \dots, t_{N-1}\}$, the population is given by
\begin{equation}
\label{Eq:W}
\vec{p}_{\text{final}} = W(t_{N-1}) \cdots W(t_1) W(t_0) \vec{p}_{\bar{n}_i}.
\end{equation}
Minimization of the final $\bar{n}$ can be used to determine an optimum sequence of pulse times.  Such an optimization procedure becomes cumbersome for larger $\bar{n}$, but generally gives rise to a series of pulses in which the majority of initial pulses have near equal pulse duration, with only the last few pulses having any significant variation.  This can be understood by looking at the structure of $W(t)$ for different pulse durations, which is illustrated in Fig.~\ref{fig:Wmatrix} for the first $50\times50$ elements.  Pulse durations are given in units of $T_f$, which is the $\pi$ time for the transition $\ket{7,0,1} \to \ket{7,-1,0}$, and the grey scale represents the effectiveness of population transfer, with a darker shade indicating a higher probability of transfer from $n_i$ to $n_f$.

\begin{figure*}[htbp]
\begin{center}
\includegraphics[width=1.0\linewidth]{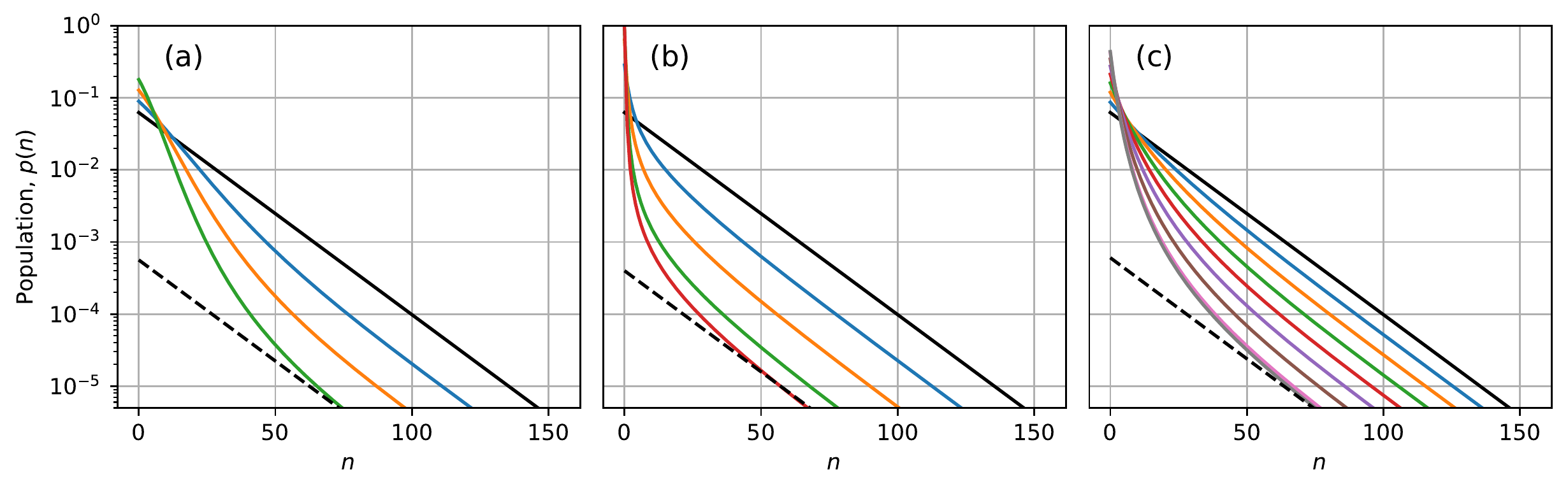}
\caption{\label{fig:combinedF7F8RSC} Evolution of the motional state distribution under a fixed DRSC pulse sequence from an initial thermal distribution with $\bar{n}=15$. 
(a) DRSC with F=7; colored lines show the distribution after every 5 pulses (15 pulses in total).
(b) DRSC with F=8; colored lines show the distribution after every 2 pulses (7 pulses in total).
(c) Two level SC; colored lines show the distribution after every 10 pulses (72 pulses in total).
The dashed lines represent a suppressed thermal distribution fit to the asymptote of the final distributions.
}
\end{center}
\end{figure*}

That shorter pulses are more effective at clearing population from the higher $n$ states can be understood from Eq.~\ref{DebyeWaller} or Eq.~\ref{DWApprox} in which the higher $n$ states have both larger coupling and less variation in coupling over a range of $n$.  This motivated us to consider a `heuristic' protocol that starts with pulses of fixed duration followed by a more tailored series of pulses near the end.  The effect of pulses of fixed duration is shown in Fig.~\ref{fig:combinedF7F8RSC} in which we plot the population distribution after a stipulated number of pulses.  For the purpose of comparison, we show results for both the $F=7$ and 8 manifolds, and traditional SC on a two level system.  In all cases we assume $\eta=0.07$ and an initial thermal state with $\bar{n}=15$, motivated by typical experimental conditions.  With populations on a log scale, a thermal state is characterized by a straight line with the slope $\ln[\bar{n}/(\bar{n}+1)]$.  In Fig.~\ref{fig:combinedF7F8RSC}, it is observed that for a wide range of high $n$ states the occupation is geometrically reduced by each successive pulse, which is evident in the equally-spaced parallel asymptotes. This suggests the population $p_N(n)$ after $N$ fixed duration pulses can be approximated as
\begin{equation}
\label{dual thermal}
p(n) = (a)^N p_{\bar{n}_i}(n) + (1-a^N) p^{(N)}_{r}(n),
\end{equation}
where $p_{\bar{n}_i}(n)$ is the initial thermal distribution, $a<1$ is a geometric suppression factor, and $p_r^{(N)}(n)$ is a residual (non-thermal) distribution which will vary from pulse to pulse.
Low $n$ states are mostly insensitive to the short duration pulses and contribute predominately to the residual distribution.  For each of the plots in Fig.~\ref{fig:combinedF7F8RSC}, the pulse duration was chosen to minimize $a$ and the total number of pulses was set to give $a^N<0.01$.  This was arbitrarily chosen to illustrate the behavior and not to suggest an optimal choice.  Indeed, one choice for two level SC would be to construct a sequence of pulses to clear the population from a large $n$ sufficient to capture say 99\% of the population.  For the $\sim 70$ pulses used in Fig.~\ref{fig:combinedF7F8RSC}(c),  99\% ground state occupation would be achieved assuming perfect transfer and neglecting any recoil heating.  However it is not obvious that such a simple pulse sequence construction is translatable to DRSC.

\begin{table}[h]
\centering
\caption{Optimum value of $a$ and the associated
pulse time $t$ for $F=7$ DRSC and $F=8$ DRSC for different initial $\bar{n}_i$.  Pulse times are scaled by the $\pi$ time for the transition $\ket{7,0,1}$ to $\ket{7,-1,0}$ for $F{=}7$ and $\ket{8,-8,1}$ to $\ket{8,-7,0}$ for $F{=}8$.}
\begin{tabular}{|c|c|c|c|c|}
\hline
& \multicolumn{2}{c|}{$F{=}7$} & \multicolumn{2}{c|}{$F{=}8$} \\
\hline
$\bar{n}_i$ &  $a$ & $t$ &  $a$ & $t$ \\
\hline
10 & 0.633 & 0.173 & 0.348 & 0.639 \\
20 & 0.787 & 0.169 & 0.577 & 0.644 \\
30 & 0.850 & 0.167 & 0.689 & 0.645 \\
40 & 0.884 & 0.166 & 0.754 & 0.645 \\
\hline
\end{tabular}
\label{tab:pulsetimeF7F8}
\end{table}

We investigate this by exploring a range of initial thermal distributions with $\bar{n}_i$ between 10 and 40 for which we determine the optimum value of $a$ and the associated pulse time for DRSC on the two different manifolds, with results tabulated in Table~\ref{tab:pulsetimeF7F8}.  The slight decrease in pulse time with $\bar{n}_i$ can be understood by the fact that $>99\%$ of population for the given $\bar{n}$ lies below the first maximum in the coupling strengths given by Eq.~\ref{DWApprox}.  The pulse time eventually increases for larger $\bar{n}$ or $\eta$.  The value of $a$ is smaller for the $F=8$ case which is expected given the larger degeneracy of states for which more quanta can be removed in a single pulse.  Less obvious is why the $F=8$ case seems more effective at removing population at lower $n$. 

\begin{figure}[htbp]
\begin{center}
\includegraphics[width=\linewidth]{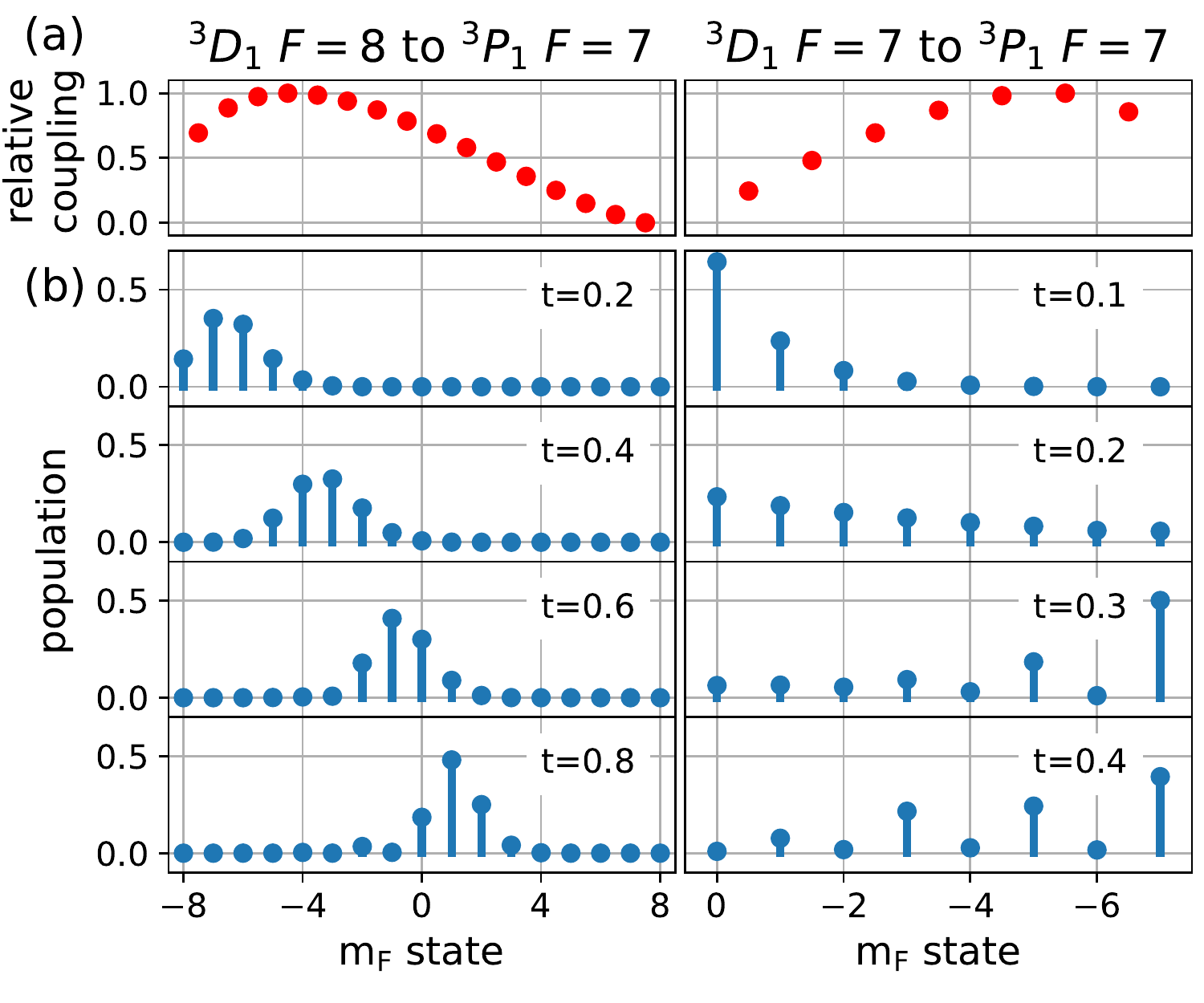}
\caption{\label{fig:Figure4} (a) Relative Raman coupling between adjacent $m_F$ states in $F=8$ and $F=7$ manifold. (b) Left (right) panels show the evolution of the pure state $\ket{-8,8,n=20}$ ($\ket{7,0,n=20}$) for a single cooling pulse in the $F=8$ ($F=7$) scheme. Pulse times $t$ are scaled by the $\pi$ time of the transition $\ket{8,-8,1}$ to $\ket{8,-7,0}$ ($\ket{7,0,1}$ to $\ket{7,-1,0}$).
}
\end{center}
\end{figure}

The difference in behavior between the $F=7$ and $F=8$ cases lies in the coupling strengths across the manifold and how this influences the transfer of population.  In Fig.~\ref{fig:Figure4}(b) we show coupling strengths from $m_F$ to $m_F-1$ for each case and the evolution of $m_F$ state population when starting from a pure Fock state with $n=20$.  As seen in Fig.~\ref{fig:Figure4}(a), in the $F=8$ manifold the coupling strength starts out large and diminishes across the manifold.  The result is that the population moves across the manifold in a pulse-like fashion enabling efficient transfer from one side to the other and removing the largest possible number of quanta.  This is crudely reminiscent of a $J_x$ operator, which would perfectly transfer population from one side to the other.  For the $F=7$ case the exact opposite occurs.  The coupling increases across the manifold which smears out the population and reduces the mean change in $m_F$ and hence the number of quanta that can be removed. 

As with traditional SC, results are likely robust to the choice of pulse sequence, provided reasonable choices are made and a sufficient number of pulses are used.  Results will be relatively insensitive to the initial pulse durations provided they are guided by Eq.~\ref{DWApprox} and steps are taken to mitigate zeros in the coupling that can occur at large $\bar{n}$ and/or increased $\eta$.  The most important consideration is the pulse lengths used in the final stages.  For conventional SC, these should eventually track the coupling strengths set by Eq.~\ref{DebyeWaller}.  For DRSC, this can be guided by the optimization of Eq.~\ref{Eq:W} for the final stages of cooling, following an initial sequence of fixed pulse lengths.  This heuristic approach is a reasonable approximation to the global optimization of all pulse lengths, at least for the cases we have explored.  In Sec.~\ref{results} we experimentally investigate the different approaches.

\section{EXPERIMENTAL SYSTEM}
Experiments are conducted in a linear Paul trap, similar in design to those reported in \cite{zhiqiang2023176lu,EMM2024}. An RF voltage at frequency $\Omega_{\mathrm{rf}} = 2\pi \times 9.385$ MHz is applied via a helical resonator. With 9 V applied to the endcaps and 0.5~V to two diagonally opposing rods, the measured secular frequencies for a single \Lu~ion are approximately $2\pi \times (0.2, 1.1, 1.2)$ MHz. The axial direction defines $\hat{\mathbf{z}}$, and a static magnetic field of 0.45 mT along $\hat{\mathbf{x}}$ sets the quantization axis.

The ion is initialized into the $^3D_1$ state via the repump beams at 350, 622, and 895\,nm. Doppler cooling and fluorescence detection are performed on the $^3D_1$ to ${}^3P_0$ transition at 646\,nm using a single beam that is linearly polarized perpendicular to the static magnetic field with three frequency components to address the $F=6$, $7$, and $8$ manifolds of $^3D_1$.  For convenience we denote these components by $D_F$. Optical pumping into the $\ket{^3D_1, F=7, m_F=0}$ state is achieved using a separate $\pi$-polarized 646\,nm beam addressing the $F=7$ to $F'=7$ transition, denoted by $D_\pi$, in conjunction with the cooling beams $D_6$ and $D_8$. Coherent operations on the $^1S_0 $ to $ {}^3D_1$ (848\,nm) and $^1S_0 $ to $ {}^3D_2$ (804\,nm) clock transitions are driven by lasers stabilized to an ultra-stable high-finesse cavity and phase locked via a frequency comb.  Microwave transitions are also used to transfer population between the $^3D_1$ $F$ states. 

State selective detection is achieved via shelving with the 848\,nm clock transitions.  Occupancy of $\ket{^3D_1, F=7, m_F=0}$ is determined by shelving population in $\ket{^3D_1, F=7, m_F=0}$ to $\ket{^1S_0, F=7, m_F=0}$ with typically 99\% fidelity, followed by fluorescence detection.  Shelving is a two-step process consisting of a microwave transition from $\ket{^3D_1, F=7, m_F=0}$ to $\ket{^3D_1, F=8, m_F=0}$ followed by an optical transition to $\ket{{}^1S_0, F=7, m_F=0}$.  Occupancy of $\ket{{}^3D_1, F=7, m_F=-7}$ is determined by directly shelving this state to $\ket{{}^1S_0, F=7, m_F=-7}$ driven by a sideband on the clock laser, which is generated by a wideband electro-optic modulator.  Fluorescence detection of population in $^3D_1$ is typically achieved with $>$99.9\% fidelity.

For DRSC, the two 646\,nm Raman beams are denoted $R_{\sigma^-}$ and $R_{\pi}$ with the subscript indicating the polarization.  These are derived from a common ECDL, detuned by $+30$~GHz from the $\ket{^3D_1, F=7}$ to $\ket{^3P_0, F'=7}$ transition. The $R_{\pi}$ beam propagates along $\hat{\mathbf{y}}$ and $R_{\sigma^-}$ along $\hat{\mathbf{x}}$. This beam configuration results in a Lamb-Dicke parameter of $\eta\approx0.07$ for the Raman transition and couples to only one radial mode of the motion.  A description of the alignment procedure and polarization characterization is given in appendix~\ref{polarization}.

As noted earlier, ac Stark shifts from the Raman beams can disrupt the degeneracy condition but this may be avoided by setting the intensity ratio $I_{\sigma^-}/I_\pi = 2$.  This leaves only the vector component from the $\sigma^-$ beam which acts as an effective magnetic field along the direction of the applied dc magnetic field.  Calibration of the beam intensities is determined via the differential ac Stark shifts on the $\ket{^3D_1,7,0} \to \ket{^3D_1,8,0}$ microwave transition, from which $\Omega$ in Eq.~\ref{DebyeWaller} can be inferred. The coupling was independently verified by observing Rabi oscillations on the $\ket{7,0,n=0} \to \ket{7,-1,n=1}$ blue sideband transition after ground-state cooling was achieved.

The Zeeman splitting of the $F=7$ level, which has a g-factor of only about 1/112, limits the strength of the carrier and hence the sideband coupling.  As a consequence, it was necessary to track the trap frequency to maintain long term resonance with the sideband.  This is accomplished using the 804 nm clock transition with an interleaved servo.

\begin{figure}[t]
\includegraphics[width=1.0\linewidth]{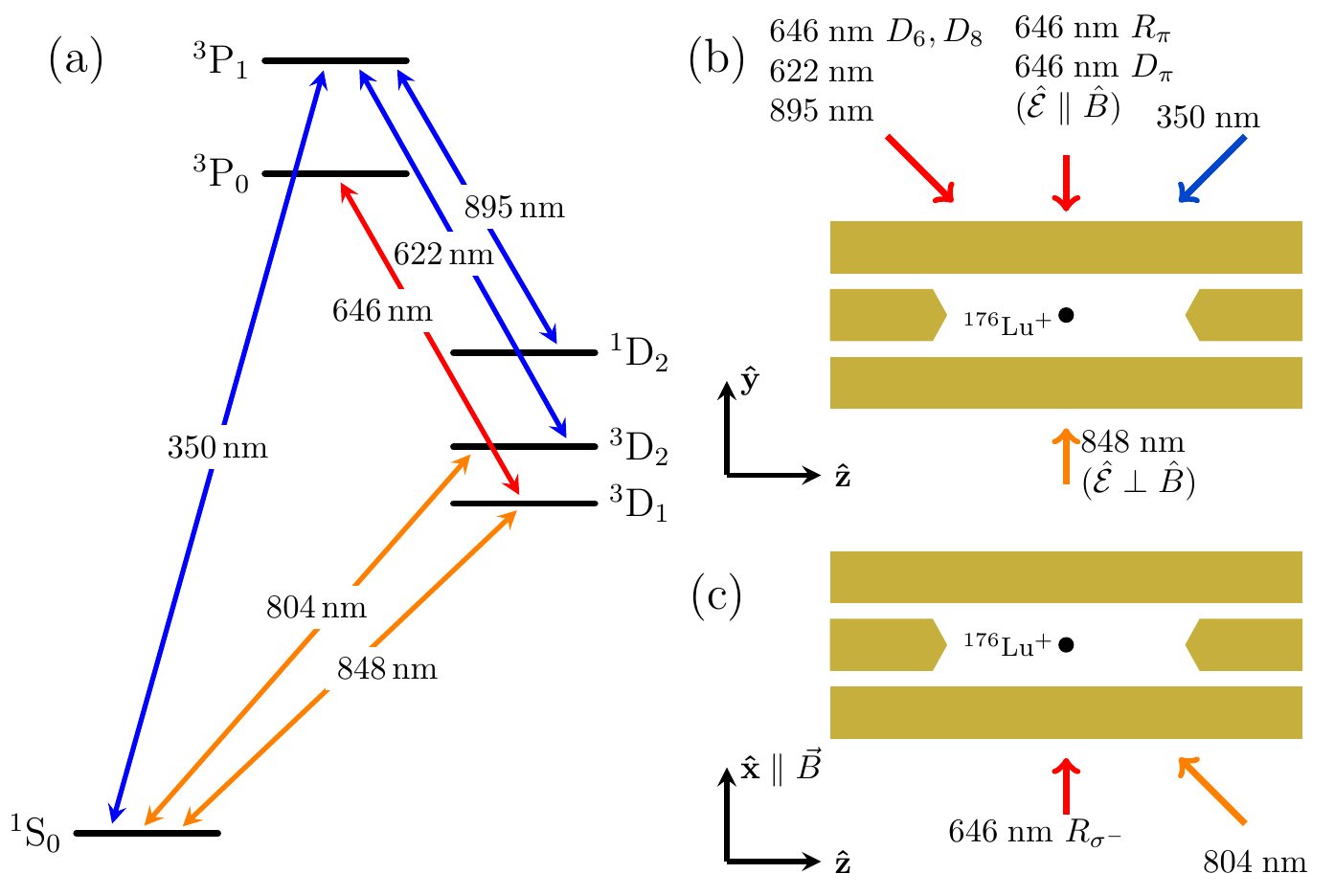} 
\caption{\label{fig:beamlevel}  (a) atomic-level structure of $^{176}$Lu$^+$ showing the wavelengths of repump (blue), cooling/detection (red), and clock (orange) transitions used. (b) and (c) ion trap geometry with laser orientations and polarizations, where relevant.}
\end{figure}

\section{EXPERIMENTAL Methods and Results}
\label{results}
The Raman sideband cooling sequence begins with Doppler cooling of a single \Lu ~ion and optical pumping into the $\ket{^3D_1, F=7, m_F=0}$ state, a Raman pulse of chosen duration redistributes population across Zeeman sublevels $m_F = 0$ to $-7$. A subsequent optical pumping pulse returns the population to the $\ket{{}^3D_1, 7, 0}$ state.

The average phonon number $\bar{n}$ is extracted using sideband thermometry on the 804\,nm clock transition. The ion is shelved to $\ket{^1S_0, 7, 0}$ as described in the previous section. We then probe a first-order sideband of the $\ket{^1S_0, 7, 0} $ to $\ket{^3D_2, 7, 0}$ transition. Success of the sideband transition is determined by reshelving $\ket{^1S_0, 7, 0}$ to $\ket{^3D_1, 8, 0}$  followed by fluorescence detection.  The ratio of the sideband amplitudes is used to estimate $\bar{n}$ by
\begin{equation}\label{eq:nsb}
\frac{P_{\text{Red}}(t)}{P_{\text{Blue}}(t)} = \frac{\bar{n}}{\bar{n}+1}
\end{equation}
which is valid for a thermal distribution~\cite{turchette2000heating}.  However, after sideband cooling the phonon distribution is no longer thermal and so the $\bar{n}$ inferred from the sideband ratio does not necessarily correspond to the true average phonon number \cite{chen2017sympathetic}.  To account for this, we simulate the cooling process to evaluated both the actual $\bar{n}$ and the value inferred from the sideband ratio, $\bar{n}_{\mathrm{SB}}$, to facilitate interpretation of the experimental results.

To further suppress the residual occupation of $n>0$ states, we implement a conditional preparation protocol, Raman dark preparation (RDP). Since imperfect Raman and optical pumping beams may scatter population out of the ground state $\ket{F=7, m_F=0, n=0}$, we apply a final Raman pulse after the last optical pumping step. This pulse removes most of the remaining $n>0$ population from $\ket{F=7, m_F=0}$. The ion is then shelved to the $\ket{^1S_0, 7, 0}$ state via the 848 nm clock transition and fluorescence detection is performed. If no signal is observed (dark state), the ion is ready for 804 nm sideband thermometry. Otherwise, the sequence is restarted from Doppler cooling. This filtering enhances the probability of preparing the ion in the motional ground state.

\begin{figure}[t]
\includegraphics[width=1.0\linewidth]{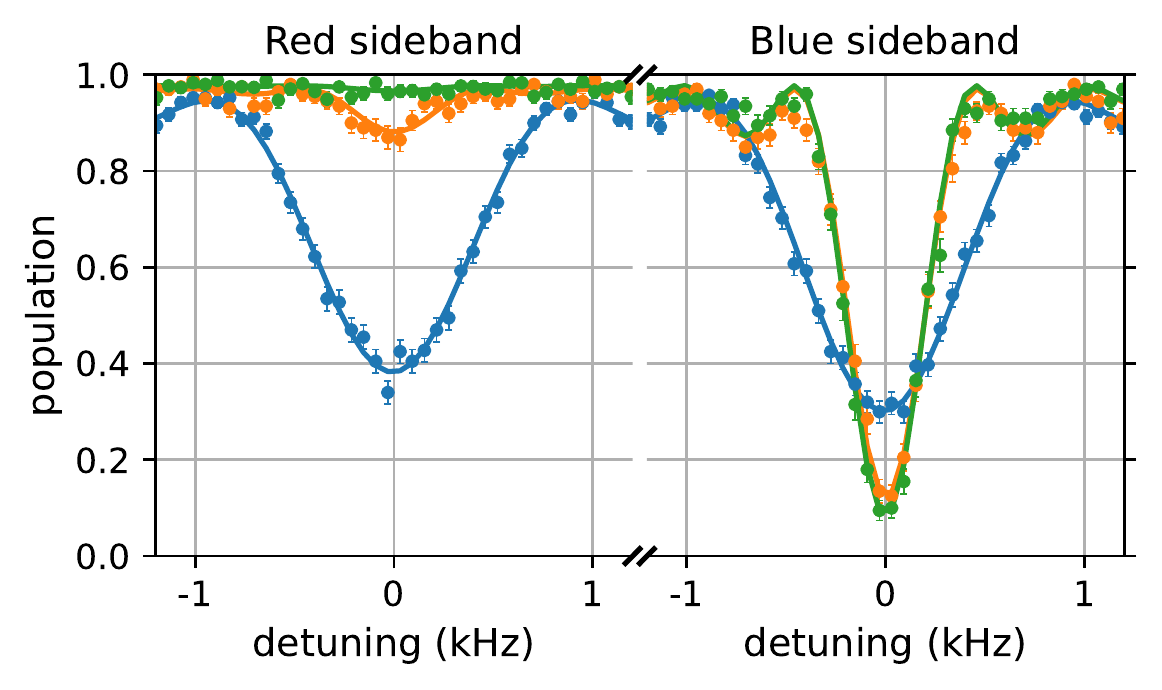}
\caption{\label{fig:804sideband} Sideband spectroscopy on the 804 nm transition after Doppler cooling (blue), 10 fixed duration pulses of DRSC (orange), and DRSC with RDP (green). }
\end{figure}

The effectiveness of the DRSC and RDP protocols is evaluated using sideband spectroscopy of the radial mode. The spectra of red and blue sidebands on the 804\,nm transition after Doppler cooling, DRSC, and DRSC plus RDP, are shown in Fig.~\ref{fig:804sideband}. Here, DRSC is implemented as 10 fixed duration pulses with the pulse length optimized to minimize the final average phonon number $\bar{n}$. From the ratio of sideband amplitudes we infer an $\bar{n}$ of $6.9(1.1)$ after Doppler cooling, which is reduced to $0.118(14)$ by DRSC, and further reduced to $0.0129(67)$ with the addition of RDP. The effectiveness of cooling demonstrated here using only a simple fixed duration pulse sequence highlights the robustness of DRSC. 

\begin{figure}[t]
\centering
\includegraphics[width=\linewidth]{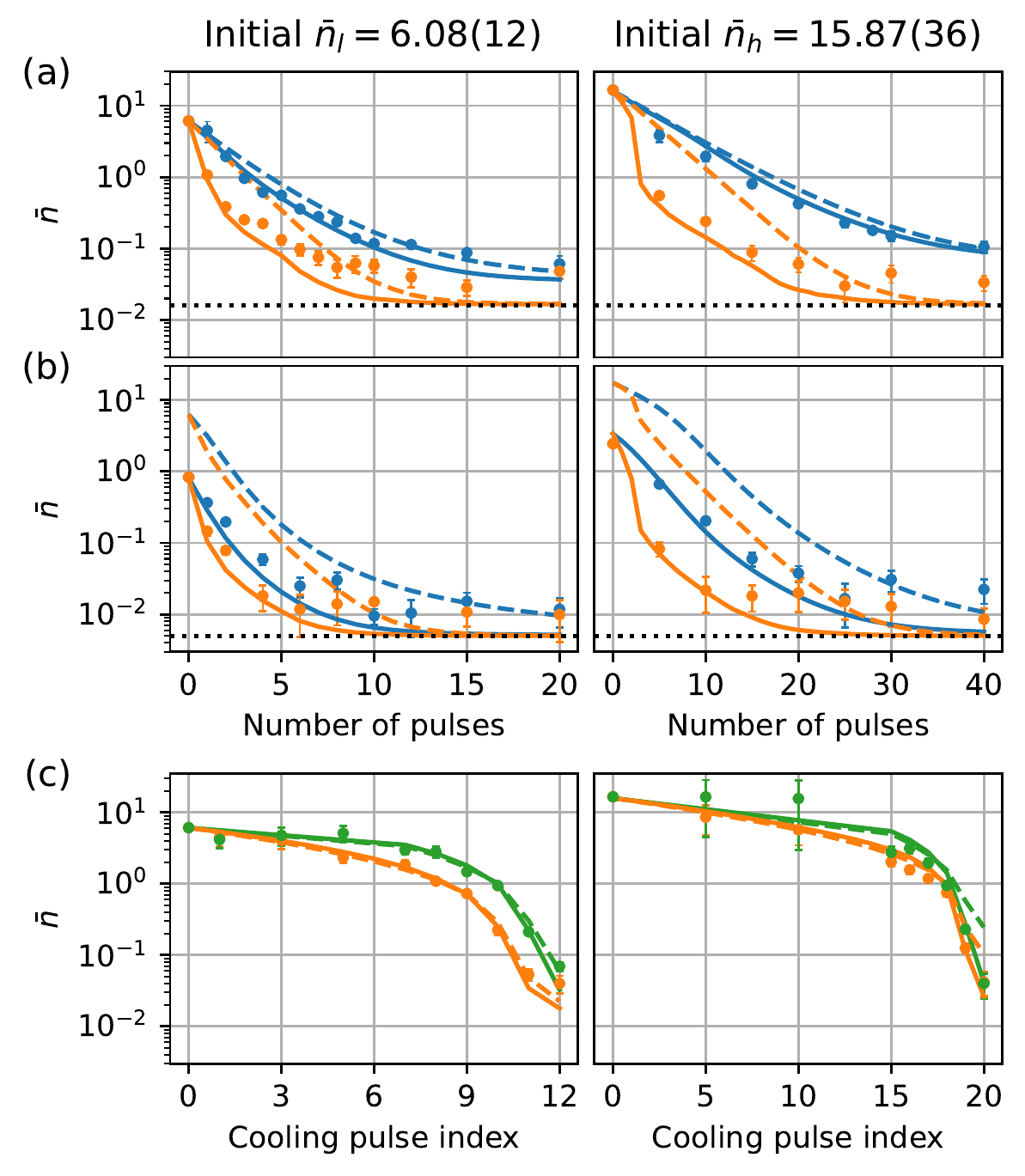}
\caption[Average phonon occupation $\bar{n}$ under various cooling protocols]{\label{fig:Ramannumber}
Average phonon occupation $\bar{n}$ as a function of the number of Raman pulses, comparing fixed (blue) and optimized (orange) cooling protocols.
(a) using DRSC only, starting from two different initial thermal states, $\bar{n}_l$ and $\bar{n}_h$, while (b) includes the additional RDP method. (c)
the evolution of $\bar{n}$ after each pulse for the heuristic approach (green) compared to the globally optimized sequence (orange), for the two initial states. 
Dashed lines indicate the simulated average phonon number $\bar{n}$, while solid lines represent the sideband-inferred value $\bar{n}_\text{SB}$. All points represent the sideband-inferred values from experimental measurements. 
}
\end{figure}

We further investigate the effectiveness of DRSC by comparing Raman pulse sequence strategies for two initial thermal states, $p_l$ and $p_h$, corresponding to $\bar{n}_l = 6.08(12)$ and $\bar{n}_h = 15.87(36)$. The higher $\bar{n}$ state was realized by deliberately detuning the Doppler cooling lasers to reduce their effectiveness. For each case, we perform a global optimization in which the pulse lengths are varied in simulation to minimize $\bar{n}$ for the given number of pulses. In Fig.~\ref{fig:Ramannumber}(a), the experimental results for the optimized DRSC pulse sequence (orange) are compared to ab initio simulations. Solid lines represent $\bar{n}_{\mathrm{SB}}$ from simulation, which may be directly compared to the measurements. Dashed lines represent the average phonon number $\bar{n}$ from simulation.  As a point of reference we also plot the evolution when using pulses of a fixed duration, which was chosen to minimize $\bar{n}$ after 12 pulses for $p_l$ and after 30 pulses for $p_h$.  All plotted simulation results include heating effects, as discussed in appendix~\ref{heating}, but these effects are excluded during the pulse optimization due to computation time.  

Results including the RDP method are shown in Fig.~\ref{fig:Ramannumber}(b).  By applying a clearing pulse after the DRSC sequence, population not in the thermal ground state is efficiently removed out of the $\ket{7,0}$ state, thereby reducing the $\bar{n}$ of the remaining $\ket{7,0}$ population which is shelved. This step also circumvents heating from the final optical pumping process resulting in the lower limit for the final $\bar{n}$ compared to the DRSC-only case.

The experimental results show good agreement with the simulations. When a sufficient number of cooling pulses are applied, the initial thermal distribution is effectively cleared. The minimum achievable $\bar{n}$ in the experiment is limited by heating from the final optical pumping pulse and the ion trap heating before the thermometry, which is detailed in Appendix~\ref{heating}. Approaching the experimental limit of the cooling, the sideband-inferred mean phonon number $\bar{n}_\mathrm{SB}$ closely approximates the true average phonon number $\bar{n}$ because the residual heating results in a final distribution which is thermal, albeit with low $\bar{n}$.

Finally, we compare the globally optimized sequence with the heuristic strategy previously outlined in Sec. II. In the heuristic approach, a fixed number of short-duration pulses is initially applied to approximately match the $\bar{n}$ of the initial thermal distribution and suppress the high-$n$ tail, such that $a^N<0.01$. This is followed by five longer pulses optimized for a thermal distribution with $\bar{n}_i = 5$, which effectively targets the remaining $p_r^{(N)}(n)$. As shown in Fig.~\ref{fig:Ramannumber}(c), the evolution of $\bar{n}$ after each pulse for the heuristic strategy shows performance close to that of the fully optimized sequence despite the significantly reduced design complexity.

\section{Conclusion}

We have demonstrated efficient ground-state cooling of a single \Lu{} ion using degenerate Raman sideband cooling (DRSC) via Zeeman-resolved transitions. By exploiting the rich internal structure of the $^3D_1$ manifold, our method enables coherent removal of multiple motional quanta within each cooling cycle, significantly reducing the number of optical repumping cycles required compared to conventional sideband cooling. 

The DRSC protocol, supported by numerical modeling and experimental validation, shows robust performance in the presence of imperfect optical pumping and recoil heating. We further introduce a heuristic cooling strategy that combines short fixed-duration pulses and a small set of optimized pulses to approach near-ground-state cooling without complex and computationally intensive global optimization.

Our simulations have been restricted to a LD parameter of $\eta=0.07$ motivated by the experimental conditions but different values would not change the essential physics.  Changes in $\eta$ will change the coupling coefficients as a function of $n$, which would result in different times for optimization, as it does for conventional sideband cooling.  In extreme cases of larger values, zeros in the coupling may arise requiring the use of second sideband transfers as noted in \cite{chen2017sympathetic}.

This approach is especially well-suited to atomic systems with large ground state degeneracy, where standard sideband cooling is hindered by inefficient repumping. The protocol can be adapted to other multi-level ions, neutral atoms or molecules, including those used in quantum logic, optical clocks, and scalable quantum computing platforms. More broadly, DRSC may serve as a general strategy to accelerate ground-state preparation in systems with complex internal structure when conventional sideband cooling becomes impractical.

\begin{acknowledgments}
This research is supported by the National Research Foundation, Singapore and A*STAR under its Quantum Engineering Programme (NRF2021-QEP2-01-P03) and through the National Quantum Office, hosted in A*STAR, under its Centre for Quantum Technologies Funding Initiative (S24Q2d0009); and the Ministry of Education, Singapore under its Academic Research Fund Tier 2 (MOE-T2EP50120-0014).
\end{acknowledgments}

\appendix

\section{Polarization analysis}
\label{polarization}
Setting and calibrating the Raman beam polarization was achieved using a combination of optical pumping and depumping at 646\,nm with a $D_\pi$ beam replacing the Raman beams.  Aligning $R_{\sigma_-}$ to the magnetic field was achieved by utilizing selection rules for the 804\,nm clock lasers as a diagnostic.  Co-propagation of these beams with the appropriate Raman laser was assured by coupling the beams through a common optical fiber.

As R$_\pi$ and $D_\pi$ differ only in the detuning, the polarization of $R_\pi$ is set by the calibration procedure used for optical pumping.  This is accomplished by optically pumping the ion into $\ket{{}^3D_1,7,0}$ and then measuring the rate at which the $D_\pi$ beam alone depumps this state.  A long depump time ensures the polarization is linear and well aligned to the magnetic field.  The measured decay time is calibrated relative to the depumping time of the $\ket{{}^3D_1,7,-1}$ transition, which is prepared by microwave transitions from $\ket{{}^3D_1,7,0}$ to $\ket{{}^3D_1,8,0}$ and then to $\ket{{}^3D_1,7,-1}$.

Once the $D_\pi$ and hence $R_\pi$ is set, the $R_{\sigma_-}$ port is aligned to the magnetic field using the coupling of the $\ket{{}^1S_0, F=7}$ to $\ket{{}^3D_2, F'=7}$ clock transition at 804\,nm, for which $\Delta m=0$ transitions are forbidden when the beam propagation direction $\hat{\mathbf{k}}$ is along the magnetic field ($\hat{\mathbf{x}}$).  For linear polarization along $\hat{\mathbf{z}}$ ($\hat{\mathbf{y}}$), the coupling on $\Delta m=0$ transitions is sensitive to rotation of $\hat{\mathbf{k}}$ about $\hat{\mathbf{y}}$ ($\hat{\mathbf{z}})$.  Suppression of the coupling on $\Delta m=0$ transitions for the two polarizations relative to the $\Delta m=\pm1$ transitions provides a measure of the alignment. In this way alignment of $\hat{\mathbf{k}}$ to the magnetic field direction to better than $0.2^\circ$ is achieved.

With the propagation direction of $R_{\sigma^-}$ set along $\hat{\mathbf{x}}$, an auxiliary $D_\pi$ beam replaces the 804\,nm beam to set the polarization.  The ion is first optically pumped to $\ket{{}^3D_1,7,-7}$ with the auxiliary $D_\pi$, $D_6$, and $D_8$ beams.  The rate at which the auxiliary $D_\pi$ beam alone depumps $\ket{{}^3D_1,7,-7}$ is then measured.  After a variable depumping time, population remaining in $\ket{{}^3D_1,7,-7}$ is shelved to $\ket{{}^1S_0,7,-7}$ to determine the remaining population in ${}^3D_1$.  The depumping time is calibrated relative to the depump time of $\ket{{}^3D_1,7,0}$. 

For calibration of the depumping times we use a resonant scattering rate for weak coupling given by
\begin{equation}
    \gamma = \frac{\Omega^2}{\Gamma} (1-p),
\end{equation}
where $\Gamma$ is the transition linewidth,   $\Omega$ is the Rabi frequency including the appropriate Clebsch–Gordan coefficient, and the parameter $p$ accounts for the probability of scattering back to the original state.  When depumping from $\ket{{}^3D_1,7,0}$ with $\sigma^\pm$, $p\approx 1/6$.  When depumping from $\ket{{}^3D_1,7,-1}$ with $\pi$, $p\sim 0.03$.   Depumping from $\ket{{}^3D_1,7,-7}$  would be predominantly from a $\sigma^+$ component, for which $p \sim 0.09$. 

From the measured depumping times for the $\sigma^-$ beam, we infer an angular misalignment of $0.15^\circ$ of $\hat{\mathbf{k}}$, which corresponds to a $0.27\%$ $\pi$ or $0.61\%$ $\sigma^+$ coupling relative to the $\sigma^-$ coupling.  For the $\pi$ beam, we get a misalignment of $0.78^\circ$, which corresponds to a relative coupling of $1.4\%$ for $\sigma^\pm$.  Any unwanted Raman couplings are therefore heavily suppressed.

\begin{figure}[h]
\includegraphics[width=1.0\linewidth]{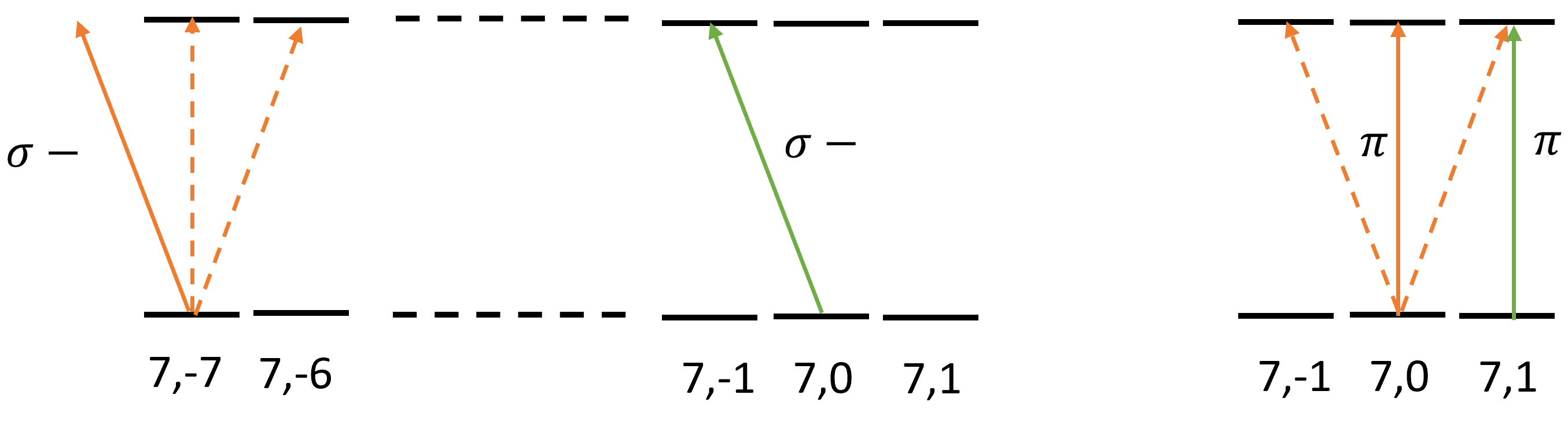}
\caption{\label{fig:depump} (Left) $\sigma^-$ beam depump procedure from $\ket{^3D_1,7,-7}$ or $\ket{^3D_1,7,0}$ and (right) $\pi$ beam depump procedure from $\ket{^3D_1,7,0}$ or $\ket{^3D_1,7,1}$.}

\end{figure}

\section{Heating evaluation}
\label{heating}
To quantify the heating associated with optical pumping, we model the repumping process as an absorbing Markov chain, with $\ket{^3D_1, 7, 0}$ as the absorbing state. Based on known branching ratios from $^3P_0$ to $^3D_1$, we compute the mean number of scattering steps required to reach $\ket{7,0}$ from all initial Zeeman sublevels. On average, 62.1 steps are needed, with $\ket{7,\pm1}$ to $\ket{7,0}$ transitions alone requiring 41.4 steps (about 0.01 quanta). 

We experimentally characterize scattering from both Raman and optical pumping processes. For Raman scattering, we insert an additional Raman pulse after the DRSC sequence and track the loss of the population from $\ket{7,0}$ as a function of pulse duration. Fitting the decay yields an effective scattering rate of 7.35(47) s$^{-1}$, consistent with estimates from the Raman $\sigma^-$ beam's off-resonant scattering rate. The scattering rate during optical pumping is determined by applying the $D_\pi$, $D_6$ and $D_8$ beams separately for varied durations and measuring the loss of population for $\ket{7,0}$, from which a total scattering rate of 41.0(2.6) s$^{-1}$ inferred.

We estimate the corresponding heating rates by multiplying the scattering rates with the mean number of scattering steps, scaled by the Lamb–Dicke parameter $\eta^2$, and incorporating geometric factors. This gives predicted heating rates of 0.8 quanta/s for Raman beams and 4 quanta/s for the optical pumping beams.

\begin{figure}[h]
\includegraphics[width=1.0\linewidth]{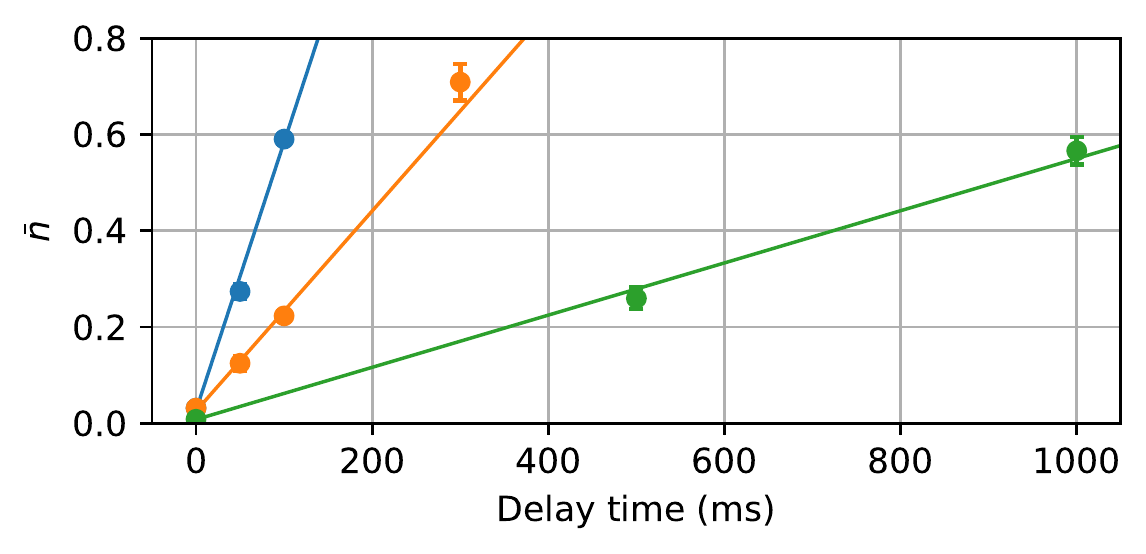}
\caption{\label{fig:heatingrate} Measured phonon number $\bar{n}$ after DRSC with (blue) additional optical pumping duration,  (orange) additional Raman duration including repumping to $\ket{^3D_1,7,0}$, and (green) addition of delay after DRSC with RDP to determine ion trap heating. Solid lines are linear fits yielding the respective heating rates.}
\end{figure}

To validate these predictions, we experimentally measure the phonon number $\bar{n}$ after applying additional Raman or optical pumping pulses of variable durations following the DRSC sequence. Results are shown in Fig.~\ref{fig:heatingrate}, and heating rates are determined from linear fits. The measured rates are $5.58(12)$ quanta/s for optical pumping and $2.078(95)$ quanta/s for Raman pulses. We find the ion trap's heating rate to be 0.553(39) quanta/s by first ground cooling and then measuring $\bar{n}$ after variable delay time.

Using the measured rates, we simulate the heating evolution via the equation ~\cite{cirac1992laser,stenholm1986semiclassical}
\begin{multline}
\frac{d}{d\tau} p_n = (n+1) A_- p_{n+1}\\
 - \left[(n+1)A_+ + n A_- \right] p_n + n A_+ p_{n-1},
\end{multline}
with mean phonon number dynamics governed by
\begin{equation}
\frac{d}{dt} \langle n \rangle = -(A_- - A_+) \langle n \rangle + A_+.
\end{equation}
Assuming equal upward and downward rates $A_+ = A_- = A$, appropriate for recoil-dominated diffusion, we construct a short-time heating transfer matrix: $H(\tau) = \{b_{ij}(\tau)\}$, with
\begin{equation}
    b_{ij}(\tau) = 
    \begin{cases}
    1 - A(2i + 1)\tau, & \text{if } j = i \\
    A(i + 1)\tau, & \text{if } j = i + 1 \\
    Ai \tau, & \text{if } j = i - 1 \\
    0, & \text{otherwise}
    \end{cases}
\end{equation}
valid for small $\tau \ll (n_\text{max}A_\text{max})^{-1} \approx 0.6$ ms given $n_\text{max} = 300$ and $A_\text{max} = 5.6$ quanta/s. For longer durations, we repeatedly apply $H(\tau)$ to propagate the phonon distribution. 

\bibliography{Raman646}

\end{document}